\documentclass[12pt,a4]{article}
\def\beq{\begin{equation}}
\def\eeq{\end{equation}}
\usepackage[dvips]{graphicx}
\begin{document}
\title{The Time Vector: An Analysis of Continuity and Change}
\author{Daniel Brown}
\newcommand{\ictitle}
{The IFE Time Vector: An Analysis of Continuity and Change to
Explain the Theories of Special Relativity and Quantum Mechanics}

\newcommand{\icauthor}
{Daniel Brown\footnote{email: daniel.brown@youmeus.com}}
\newcommand{\icaddress}
{University College London, U.K.}
\begin{titlepage}
\begin{center}
{\large{\bf\ictitle}}
\bigskip \\ \icauthor \\ \mbox{} {\it \icaddress} \\ \today \\ \vspace{.5in}
\end{center}
\setcounter{page}{0}
\begin{abstract}
This paper sets out to explain: \\
1. Why the speed of light c is a constant and is the maximum speed
at which any moving entity can travel. \\
2. Why time elapsed differs for a moving entity relative to a
stationary entity. \\
3. Reasons for the confusion between the wave and particle
nature of an entity.\\
4. The relation between speed of light c, Planck's constant k and time.\\
5. The probability for a freely moving entity interacting in a particular spatial position.\\
6.  Expressions for Mass and Momentum using this notation.\\
7. The minimum locus of uncertainty in position and momentum.\\

\end{abstract}
\end{titlepage}
\section{Identity}

The conceptual apparatus for investigating time, space and energy
and the concept of identity require revision when analysing change
at the microscopic level. \\

We make use of the notion of change all the time. A cup of tea
falls on the floor and it breaks. A broken cup results. A child
transforms over time into an adult. A train moves along its track.
A lump of metal is beaten into a ring.\\

Yet change is not a simple concept. It integrates interacting
phenomena and concepts whose relations become progressively more
difficult to define at increasingly fine
levels of granularity. \\

Analysis of change requires a very precise reference of the
identity of the thing that is changing. Paradoxically the net
which precisely captures this concept must cover both that which
was, together with that which is somehow different. Lao-tzu
highlighted the difficulty with his observation: ``if you realise
that all things change, there is nothing you will try to hold onto''. \\

We require elucidation both of notions of continuity and of
identity. Understanding what it means for a thing to remain the
same clarifies what it means for that thing to change. \\

Logically we start with a theoretical examination of the smallest
element of invariance possible - how a thing remains the same from
moment to moment. Consider the smallest possible theoretically
adjacent moments (the concept of adjacency is significant in
understanding change) of the smallest
possible thing. \\

To illustrate an example in everyday parlance, we can assume that
when a teacup is hurled towards the floor, then at the points when
it is flying through the air, it remains the same teacup. Once it
lands and splinters to pieces on the floor, then we say that it
changes into shards of pottery. If glued back together, then we
describe it as ``pretty
much'' the same teacup as at the outset. \\

We might, however, be more cautious in observation. Firstly, when
the teacup accelerated and changed its speed (i.e. when launched
and hurled) we could (pedantically) say that the teacup at that
point changed from one entity into another. - Of course, we
typically assert that the teacup did not \textit{fundamentally}
change, but that it was acted upon in such a way that it remained
the same \textit{thing} and that its \textit{properties} changed -
this is how we usually utilise the concept of identity to enable a
definition of state change. However, at the microscopic level, it
may not be possible to retain this separated concept of identity
and associated
properties. \\

We shall therefore include the concepts of space and even time in
our most theoretically precise definition of ``thingness'' and
proceed to examine more closely the concept of invariance. \\

Consider a hypothetical stationary teacup (a theoretical unit of
identity teacup that has no set of jiggling sub-components etc.).
We observe this teacup closely from one moment $(t_1)$ to the next
$(t_2)$ and are absolutely satisfied that this teacup has not
changed in any way. - Specifically, we require that there has been
a change in the time of the viewer (which implies that there has
been a change in state of the viewer) but does not imply that
there has been a change in time experienced by the teacup. Indeed,
if nothing at all has changed in the teacup then we assert that
time will not have progressed from the point of view of the
teacup, and this, we
assert, defines our most stringent notion of invariance. \\

We \textit{define} the teacup as the unmoving matter at place
$x_1$ and at moment $t_1$. We can use a Space, Energy, Time
(S,E,T) notation to describe the teacup: $(x_1,e_1,t_1)$. \\

We make the following Assumptions:\\
(i) time, space and energy advance in quantised units\\
(ii) time can only advance when change occurs\\
(iii) change can only occur if there is either or both:\\
 \hspace* {10mm}(a) change in energy\\
\hspace* {10mm}(b) change in spatial position \\

If an entity changes spatial position or changes energy, then we
assume that each of these changes takes a quantity of time which
we shall define as $t^{*}$ and t{'} respectively (later we shall
see that  these components can be combined as vectors on two
orthogonal axes and it is the purpose of this paper to explain the
very great significance of the combination of these two elements
for the advance of time). Then we have $(x_2,e_1,t_{2}^*)$ or
$(x_1,e_2,t_{2}')$ (here, as elsewhere we shall use $t^{*}$ and
t{'} to refer to the smallest possible units of change in
space-time and energy-time respectively) and we typically explain
that it is the matter but not its position or the time associated
with that matter that constitutes its identity. Indeed, we take
this so much for granted that the last clause of this phrase
sounds very awkward. However, at the microscopic level, these
features cannot
be considered as we might traditionally expect. \\

At the microscopic level, we shall consider an entity through all
three elements of its matter, its position in space, and its
position in time. We can then explain (partial) continuity of
identity in different ways - for example: (i) continuity of energy
over a changing arena of space and time.
(ii) continuity of space over a changing arena of energy and time\\

This complexity in identity definition is highlighted in the
macroscopic world in the paradox of the ship of Theseus where over
a period of time, in order to repair a wooden ship - which we can
refer to as Ship 1 - its planks are replaced one by one. However,
the original planks are taken aside and reconstituted in identical
architecture into another ship at a different location which we
call Ship 2. The question of identity is which ship is the
original ship of Theseus. - In order to decide, we need to decide
on our criterion of identity: either it is continuity of matter
over changing space and time (Ship 2) or continuity of space over
changing matter and time (Ship 1). In a very real sense, both
ships represent two parallel continuities of identity over
changing time. This forms a useful analogy for this paper as it
relates to the potential for an object to be conceived  as in two
alternative
positions at the same time.\\

How we regard the identity of an entity can therefore affect both
what and where we presume that entity to be. In particular, it is
possible to view an entity as being at two different points in
space at the same time, dependent on how we have tracked and how
we collapse its identity. \\

The rest of this paper will be concerned with the different
implications for a change in time caused by a change in spatial
position and a change in time caused by a change in energy, and
the consequences that this dual (vector) approach to time has in
terms of relativity and quantum mechanics.

\section{Interrelating Fluctuating Entities}

In order to analyse more precisely the mechanics of a vector
approach to time, it is necessary to formalise a set of rules
governing the interaction of time, space and matter. A simple
example in the macroscopic world is what has recently been termed
the ``Mexican wave" in football stadiums. This illustration is
important philosophically because we can examine the contribution
of each individual fan and visualise subtle interactions at a
theoretical level.\\

We shall explore the mechanism by which fans propagate a wave to
undulate across the stadium. - We assume first of all that the
stadium has only two rows A and B. Row A comprises n adjacent
fans. In row B we are only concerned with two fans: one at the
start of the row and the second at the nth position. Each fan can
move through a cycle of discrete states of standing up and sitting
down as illustrated in diagram 2.\footnote{the lateral effects of
entities on each other are significant; however in this
paper we shall examine just one spatial dimension.}: \\

DIAGRAM 1 \\

Row A $\qquad \triangle\triangle\triangle\triangle\triangle\triangle\triangle\triangle\triangle\triangle\triangle\triangle\triangle\triangle\triangle\rightarrow$ moving fan event \\

Row B $\qquad \triangle$  fan1 $\qquad \qquad \qquad \qquad \quad
\triangle$ fan n\\

The two fans in Row B measure time elapsed whilst remaining
stationary (these are simply individual fans bouncing up and down
and not interacting with other fans) and counting energy changes.
Time elapsed is also measured by the ``fan event'' in Row A
travelling from the position of fan 1 to the position of fan n. We
shall examine carefully how each fan measures passing time.\\

The following rules enable us to specify - and vary - the
propagation of a wave around the stadium: \\

DIAGRAM 2\\
 (Different Energy states represented by a fan standing up and sitting down)\\
\hspace* {70mm} O
\\[-4mm]
\hspace* {35mm} O \hspace* {28mm}H \hspace* {31mm} O
\\[-4mm]
 \hspace* {1mm} O \hspace* {28mm}H \hspace* {31mm}H \hspace* {30mm}H \hspace*
 {32.5mm}O
\\[-4mm]
H \hspace* {30mm}H \hspace* {31mm}H \hspace* {30mm}H \hspace*
{31mm}H
\\[-4mm]
H \hspace* {30mm}H \hspace* {31mm}H \hspace* {30mm}H \hspace*
{31mm}H
\\[-4mm]
H \hspace* {30mm}H \hspace* {31mm}H \hspace* {30mm}H \hspace*
{31mm}H
\\[-4mm]
HHH \hspace* {26mm}H \hspace* {29mm}H \hspace* {32mm}H \hspace*
{29mm}HHH
\\[-4mm]
\hspace* {6mm}H \hspace* {28mm}H \hspace* {27.5mm}H\hspace*
{35mm}H \hspace* {33mm}H
\\[-4mm]
\hspace* {6mm}H \hspace* {28mm}H \hspace* {27.5mm}H\hspace*
{35mm}H \hspace* {33mm}H
\\[-4mm]
\hspace* {6mm} HH \hspace* {26mm}HH \hspace* {23mm}HH \hspace*
{32mm}HH \hspace* {28.5mm}HH
\\[-4mm]
\\[-4mm]
 \hspace* {1mm} E=0 \hspace* {24mm}E=1 \hspace* {26mm}E=2 \hspace* {25mm}E=1 \hspace*
 {26mm}E=0

1.  All fans move in exactly the same way\\

2. Each fan has a certain number of discrete states q. We will
call these ``energy states'' e.g. in Diagram 2 there are 3
discrete energy states 0,1 or 2. Each fan is capable of
registering states of energy, a quality which we annotate by e.
We indicate the first energy position by 1, the next by 2$\ldots$100 etc.\\

3. A fan can have an effect on an adjacent fan only.\\

4. A fan is activated when its adjacent neighbour moves from one
specified energy state to another - the ``trigger point'' p. For
example when a fan moves from 1 $\rightarrow$ 2. This trigger
point can be varied. We could alternatively specify that the
downward motion of moving from 2 $\rightarrow$ 1 initiates an
upward motion of 0 $\rightarrow$
1 in an adjacent fan. A fan will not commence moving until its adjacent fan passes through the trigger point.\\

5. Once a fan is acted upon, this fan will move through a complete
cycle of energy states (e.g. Energy moves through 0,1,2,1,0 in the
diagram above).\\

6. Each fan can \textit{\textbf{itself measure time only through a
change in energy-state}}. Each fan records time by holding up a
card which indicates its registered time for which 1 indicates the
first unit of time, 2 the next $\ldots$100 etc. Thus the amount of
time recorded by a fan to reach its trigger point at any
spatial position is $p t'$.\\

7. Each change of energy state causes a change t' in local time
where local time applies to the current fan. \\

8. Each fan sits a distance d apart from each other fan. A spatial
variable S indicates the position of each fan: 1 indicates the
position of the first fan, 2 that of the next fan $\ldots$100 the
$100^{th}$ fan etc. Hence if we move from the first fan to the nth
fan, the distance travelled is $n(-1) d$. For convenience we can
refer to a fan at the 0th position and hence distance travelled is
nd.\\

9.  Each activation of a neighbour is instantaneously associated
with a change in non local time $t^{*}$ for  adjacent fans (but
not the moving fan itself) - this is effectively the time taken
for the energy movement to be propagated from one location to its
adjacent position. Significantly it appears empirically that
generally $t^*
> t'$ and that $pt' \gg t^*$\\

10. The unit change in time t* caused by a change in space is
distinct from the unit change in time t' caused by the change
in energy.\\

11.  The final position of a fan is determined by an
\textit{interaction}. This occurs at a specific spatial position
and energy. Once an interaction has occurred, then all activities
of that fan event cease at that time.\\

There cannot logically be a change in time without a change in
either energy or space. Causally, therefore, for a given entity in
a specific fixed spatial position, then with no change in energy
\textit{there can be no change in
time}.\\

We now have a method for indicating the position of state of each
fan at each position according to (S,E,T) which defines a position
in space, a state of energy and a position in time. The
combination of rules with the method of representation enables us
to abstract from the fans themselves to examine the consequent
dynamic interacting fluctuating entities that arise.\\

We indicate a change of energy by (e), a change of spatial
position by (s) and a change in time by (t). Likewise \textbf{e},
\textbf{s}, and \textbf{t} indicate that these quantities of
energy, space and time respectively remain the same. Since they
are in a cycle of interaction, it is not appropriate to attribute
the concept of ``thingness" to time, space or energy in isolation
and we shall therefore refer to these virtual concepts as
entities only.\\

We can then extract a set of basic Interrelated Fluctuating Entity
- IFE - Rules (``iffy" conveniently catches the component temporal vagueness) which determine a movement of the disturbance through space, energy and time:\\

\textbf{IFE Causal Rules}

(i) A change in energy (e) at a certain position in space and time
causes an (instantaneous) increment in time (t') associated with
the new energy state at the same position in space \textbf{s}:

 \textbf{(e) \textbf{s}$ \rightarrow (t')$      a change in energy over
constant space causes a change in time\\}

(ii) A change in time at a certain position in space associated
with a certain energy causes a change in energy (e) at the same
position in space \textbf{s} - at the later time. Rules (i) and
(ii) will continue in oscillation at the same space for p
oscillations until the trigger point is reached. It is useful to
consider a simple instance for a single movement in space where
p=1, i.e. where a single movement in energy is sufficient to
result in a movement in space.

(where $p > 1$ then we can consider the cycling of oscillations
between increments in time and energy as taking (p-1) t' following
which the energy reaches the trigger point so that there is
effectively a movement
in space once there have been p energy increments).\\

(iii) If the change in energy (e) at a certain position in space
and time passes the trigger point p, then this causes an increment
in space (s) associated with the new energy state at the same
position in time t:

 \textbf{(e) \textbf{t}$ \rightarrow (s)$      a change in energy over
constant time causes a change in space\\}

(iv) The change in space (s) at a specific time of a specific
energy causes an instantaneous increment in time (t*) associated
with the energy state at the new position in space:

 \textbf{(s) \textbf{e}$ \rightarrow (t^*)$ a change in space over
constant energy causes a change in time}\\

These IFE rules are sufficient to define a disturbance which moves
with a constant velocity through space and time. It will be noted
that the disturbance has inertia and will continue to move
indefinitely with this constant velocity
(until it interacts with another entity).\\

At the trigger point a change in (e) results \textbf{both} in a
change in energy-time (rt') \textbf{and} - because the trigger
point has been passed - a change in space (s). The change in (s)
causes a change in space-time $(nt^*)$. The total magnitude of
time elapsed must then comprise both (rt') and $(nt^*)$.

We can (crudely) illustrate in two dimensions the causal cycle:\\

DIAGRAM 3\\
\\[-4mm]
$(e)
\longrightarrow\longrightarrow\longrightarrow\longrightarrow\longrightarrow
(s)
\\[-4mm]
\hspace* {-.5mm} \uparrow \hspace* {1mm}\downarrow \hspace* {30mm}
\swarrow
\\[-4mm]
\hspace* {1mm}\uparrow\hspace* {.8mm} \downarrow\hspace* {26mm}
\swarrow
\\[-4mm]
\hspace* {1mm}\uparrow \hspace* {.8mm}\downarrow (t*) \hspace*
{13mm} \swarrow
\\[-4mm]
\hspace* {1mm}\uparrow \hspace* {.8mm}\downarrow \hspace* {18mm}
\swarrow
\\[-4mm]
\hspace* {1mm}\uparrow \hspace* {.8mm}\downarrow\hspace* {14mm}
\swarrow (t')
\\[-4mm]
\hspace* {0mm}\uparrow\hspace* {.8mm}\downarrow\hspace* {10mm}
\swarrow
\\[-4mm]
\hspace* {1mm}\uparrow\hspace* {.8mm}\downarrow\hspace* {6mm}
\swarrow
\\[-4mm]
\hspace* {1mm}\uparrow\hspace* {.8mm}\downarrow\hspace* {2mm}
\swarrow
\\[-4mm]
(T)$
\\

We note some important consequences. Firstly, it is the occurrence
- at the trigger point - of both a movement in space and its
associated movement in time, and the movement in time associated
with this change in energy which results in a fundamental
ambiguity for a given magnitude of time - where an entity is
located in space and what its energy is. Secondly it is the causal
link between a movement in energy, a movement in space and an
associated increase in time which is responsible for the maximum
speed of light. Thirdly there is a possibility for a very small
single movement in energy and space without an associated movement
in time which has potential implications for the collapse of an
IFE which would in principle permit particular effects over a
distance (produced specifically by a ``domino" effect of single
movement rippling across adjacent spatial positions) at a speed
greater than light.

The consequence is that from a temporal perspective there has been
both an advance in energy-time $t'$ and an advance in spatial-time
$t^{*}$. We must then resolve these separate but coterminous
advances in time which commence from the same initial moment into
a single magnitude.

Analysing these dual advances in time is not at first natural to
our way of thinking. We have to consider that \textbf{as the
changes in time t' and t* occur, they operate from the same start
point instantaneously: the increments apply in both instances to
$t_1$ and whilst t* is logically subsequent to t', it is not
temporally subsequent: \textit{t* and t' therefore apply in unison from the same moment}}.\\

An alternative way of viewing this is consider that time advances
in two ways at once: on the one hand through a movement in energy
and on the other through a movement in space. And both occur
coterminously.\\

We define the total time state or time magnitude of an entity at a
certain spatial position and energy to include both the time t'
taken for changes in energy position and the time $t^*$ taken for
changes in spatial position. ``Experienced'' time, which depends
solely on t', may vary across a distance between a moving entity
and a stationary one, but \textbf{all interactions occur between
entities which are in the same time
state}.\\

Let us examine the time state for adjacent observers and a moving
entity:\\
\\
\\
DIAGRAM 4 \\
(Row A) $\qquad \triangle\triangle\triangle\triangle \rightarrow$
moving fan \\ (Row B) $\qquad \triangle_1\triangle_2\qquad$ (1,2)
indicates position number
\\

We can compare the time \textit{experienced} by the spatially
moving entity for it to move in Row A from adjacent point 1 to
point 2, and the \textit{time
state} of this spatially moving entity when it reaches position 2.\\

For the spatially moving entity, the time \textit{experienced} is
simply pt'. However, the time state of this entity, which
corresponds
with the time \textit{at which an interaction can occur} will be different.\\

The complete time state of the moving entity which will be
observed by the adjacent fans in row B must include the time for
spatial movement (which will not have been \textit{experienced} by
the fan) $t^*$.

From Diagram 3, we note that:\\

(i) the increase in time t' which is caused by the change in
energy applies from the initial time $t_1$\\

(ii) the increase in time t* which is caused by the change in
space
also applies from the initial time $t_1$\\

\textbf{Both increases in time t' and t* operate at once, at ``the
same time''}. In order to calculate the time state: we cannot
simply add t' and t* for these increments are operating
simultaneously and originate from the same initial time $t_1$. \\

These advances in time operate in unison, and together they define
the quantity of advance in time. \textbf{In order to combine these
coterminous advances in time which proceed along different axes of
energy-time t' and space-time t* into a single time magnitude, we
shall make the following hypothesis: that these axes are
orthogonal and hence that their combination comprises a simple
pythagorean sum. It is remarkable that the consequence of these
two orthogonal dimensions of time is both the theories of special
relativity and quantum mechanics as will be shown.}

Let us consider what occurs in a time of magnitude $|T|$ from a
start time $t_1$

$$|T| = t_1+\sqrt{(t^*)^2+(t')^2}$$

Where the start time $t_1 = 0$ then
$$|T| = \sqrt{(t^*)^2+(t')^2}$$

In the case where the trigger point p is greater than 1 then:

$$|T| = \sqrt{(t^*)^2+(pt')^2}$$

And for a sequence of n spatial movements:

$$|T| = n\sqrt{(t^*)^2+(pt')^2}= \sqrt{(nt^*)^2+(npt')^2}$$

Following a series of n spatial movements, in the final nth
spatial position there follows a sequence of variable r energy
movements (which may exceed the trigger point i.e. we can have
$r>p$). $rt'$ corresponds to the actual \textbf{detected} energy
of the IFE at the point of its interaction where energy $e =
\frac{h}{rt'}$. We will refer to $\frac{1}{rt'}$ as the
``frequency" of the IFE, since this mathematically corresponds
with the terminology used in the literature. However, it will be
seen that the the rt' term is somewhat different conceptually.
Since rt' is measured energy-time t' then we can form the complete
expression for the time magnitude:

\beq |T| = \sqrt{(nt^*)^2+(npt'+rt')^2} \eeq

\textbf{Since n and r are variables, there exists a range of
alternative combinations of energy and spatial position states
which combine to form the same time magnitude $|T|$ from variable
components of energy-time t' and space-time $t^{*}$. We can
represent this for a fixed $|T|$ of magnitude qt' as a ``temporal arc" (see as in diagram 5 below}):\\

\pagebreak

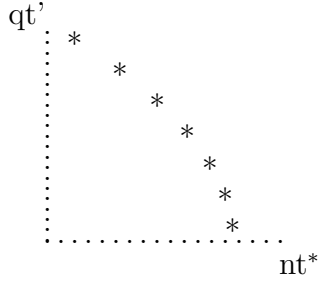
Diagram 5 - temporal arc for a photon at a time $|rt'|$ \\
\hspace* {-6mm} qt'
\\[-4mm]
 \vdots\hspace* {2mm}*
\\[-4mm]
 \vdots\hspace* {8mm}*
\\[-4mm]
\vdots\hspace* {13mm}*
\\[-4mm]
\vdots\hspace* {17mm}*
\\[-4mm]
\vdots\hspace* {20mm}*
\\[-4mm]
\vdots\hspace* {22mm}*
\\[-4mm]
\vdots\hspace* {23mm}*
\\[-8mm]
\ldots\ldots\ldots\ldots\ldots\ldots \hspace* {23mm}
\\[-4mm]
 \hspace* {30mm}
nt$^{*}$\\

\textbf{All points on the temporal arc have the same time
magnitude.}

It is notable that we can represent the time state as a complex
vector. Thus using a notation with spatial-time ($t^*$) as real
and energy-time (t') as imaginary: \beq \b{T} = n t^* + \imath (n
p + r) t' \eeq or where $z =( p + r/n)$ : \beq \b{T} = n (t^* +
\imath z t') \eeq

Differences in experienced time between moving and stationary
entities all stem from the difficulty in measuring $t^*$.

\section{Properties of Interrelating Fluctuating Entities}

We can now define some properties of interrelating fluctuating
entities:\\

1. the time to travel from one IFE to the next is a complex number
(or equivalently a two dimensional vector) where
$$ \b{T} = n t^{*}  +   \imath(n p + r) t' $$
or where  z = (p + r/n) :\\
 $$\b{T} = n (t^{*} + \imath z t') $$

 Time Magnitude $ {\mid}  \b {T}  {\mid} = (T) = \sqrt{
{T}{T^*}} = \sqrt{(nt^{*})^{2} + (npt'+rt')^{2}} $

For large n, we can often ignore the residual rt' energy-time
component in calculations of time magnitude. For increasingly
small distances, however, it becomes increasingly significant.\\

2. Frequency f = $\frac{1}{\imath (rt')}$\\

3. The speed of propagation

\beq  v = \frac{nd}{|T|} = \frac{nd}{\sqrt{(nt^*)^2+(npt'+rt')^2}}
\eeq
\\

4. Wavelength $ \lambda =   v/f = \imath qt'v$

\beq  \lambda =
\frac{\imath(nd)(qt')}{\sqrt{(nt^*)^2+(npt'+rt')^2}} \eeq
\\

5. A consequent maximum speed is implied for which a wave can
theoretically propagate through the medium. This will occur when
the trigger point p is zero. i.e.

\beq v_{max} = \frac{d}{\sqrt{(t^*)^2+(0)^2}} = \frac{d}{t^*} \eeq

This is significant as it represents purely the time taken to move
spatial distances by an entity where no energy changes are
occurring. $v_{max}$ is the speed of light c, and the total
absence of energy-time as a component in the time magnitude
explains why such a speed cannot be exceeded and why for an entity
travelling at such a speed, we would expect no time to be
experienced by that entity at all.

\section{The Magnitude of the Time Vector (over large distances)}

From the Time vector equation:

T = n (t*  +  $\imath$ z t')       where z = (p + r/n)

We can calculate the magnitude of this time vector as:

\beq |T |= n \sqrt{(t^*)^2 +(zt')^2} \eeq

In a sense this simple equation is all we need to express the
theory of relativity, for $|T|$ expresses the total time magnitude
and (zt') represents the time experienced by the moving IFE. In
order to demonstrate how this accords with the familiar model of
the theory of relativity, we can further calculate:

$$ \textmd{Speed } v = \frac{nd}{n \sqrt{(t^*)^2 +(zt')^2}} $$

\beq v = \frac{d}{\sqrt{(t^*)^2 +(zt')^2}} \eeq

 we can assert also that for the speed of a photon over a
significant distance  there is no trigger point (i.e. p=0) and q/n
will be comparatively very small then: \beq \textmd{Speed } c =
\frac{nd}{nt^*} = \frac{d}{t^*} \eeq

Rearranging  (7):

$$ |T| = n\left( \frac{(t^*)^2}{\sqrt{(t^*)^2 +(zt')^2}} +
\frac{(zt')^2}{\sqrt{(t^*)^2 +(zt')^2}}\right) $$

Substituting from (8) and (9) into the first part of the
expression and rearranging the second part:

$$ |T| = \frac{nv(t^*)}{c} + n(t^*
zt')\frac{(zt')}{t^*}\sqrt{\frac{1}{(t^*)^2 + (zt')^2}} $$

Further rearranging the second part:

$$ |T| = \frac{nv(t^*)}{c} + n(t^* zt')\sqrt{\frac{(t^*)^2 +
(zt')^2 - (t^*)^2}{(t^*)^2[(t^*)^2+(zt')^2]}} $$

From which we obtain:

\beq |T| = \frac{nv(t^*)}{c} + n(t^* zt')\sqrt{\frac{1}{(t^*)^2} -
\frac{1}{(t^*)^2 + (zt')^2}} \eeq

But from (8) and (9) we have:

\beq \frac{\sqrt{c^2 - v^2}}{c} = t^* \sqrt{\frac{1}{(t^*)^2} -
\frac{1}{(t^*)^2 + (zt')^2}} \eeq

Substituting this expression into (10) we obtain:

\beq\ |T| = \frac{nv(t^*)}{c} + n \frac{\sqrt{c^2 - v^2}}{c} (zt')
\eeq

Now in terms of distance travelled x:
$$x =  c(nt*)$$

Substituting into (12) we arrive at:

$$ |T| = n(zt')\sqrt{1-v^2 / c^2} + (v / c^2)x $$

Since n(zt') corresponds to $\tau$ the amount of time experienced
from the perspective of the moving entity and $|T|$ corresponds to
the time observed by a stationary observer, this is the familiar
Einstein-Lorentz expression:

\beq\tau = \gamma(|T | -  (v x/ c^2 ))   \textnormal{   where }
\gamma = (1 - v^2 /c ^2)^{-1/2} \eeq

The simplicity and explanatory power of the vector approach is
notable by comparison. Even in cases (such as in the calculation
of combined velocities later in this paper) where there may be an
eruption of terms, there is a fundamentally comprehensible
approach, which is not always the case with Einstein-Lorentz
presentations.

It is important to recognise that all our understanding of
``relativistic" effects are fundamentally underpinned by time and
time alone. Calculations for alterations in distance arise from
the perception of measured space through velocities which
ultimately refer back to differences in experienced time derived
from the difference between combination of space-time and
energy-time.

\section{Planck's constant and the speed of light c}

It is mathematically trivial but philosophically interesting to
examine the constants that underlie the units of the axes of
space-time t* - in terms of distance, and energy-time t' - in
terms of energy.

We know that Distance = Speed . Time:

Then
 \beq\ x = n c t^{*} \hspace* {4mm}\textnormal{ i.e.  } \Delta x= ct^{*}\eeq

Where c, the speed of light is revealed as the constant that
connects the smallest possible quantum spatial distance d =
$\Delta x$ to the smallest quantum of spatial-time $t^{*}$.

It has been observed empirically that Energy = h. Frequency:

\beq  e = \frac{h}{\imath(rt')}\hspace* {4mm}\textnormal{ i.e.
}\Delta e= \frac{h}{\imath(t')} \eeq

Where h, Planck's constant is revealed as the constant that
connects the smallest possible quantum in energy $\Delta e$ to the
smallest quantum of energy-time ${t'}$.

This refines our view on the meaning of the constants c and h and
the connections from time to distance and energy respectively.

\section{The probability for a freely moving entity interacting in a particular spatial position}

The magnitude of the time vector associated with the notion of the
temporal arc indicates that at a specified time magnitude, there
are variable combinations of spatial position and energy which can
combine in an IFE to form this same time magnitude. For small
distances, the contribution of the energy-time component $\imath
rt'$ becomes increasingly significant and when the total time
magnitude $|T|$ measured is of the order of $ |\imath rt'|$ then
there will be challenges in precisely divining specific energy and
spatial position. Indeed it is immediately evident that for a
given $|T|$ there is unavoidably some uncertainty in defining
these qualities, and that this uncertainty will become
increasingly pronounced for smaller values of $|T|$. \textbf{It is
because there are alternative compositions of spatial-time (nt*)
and energy-time (rt') for a given time magnitude $|T|$ that only a
statistical method can be used to define the position in space and
energy of the IFE}.

Calculation of P(x) the probability of the IFE being located
(through an interaction) in a specific position is somewhat more
intricate than might at first be expected.

Consider an IFE starting from an initial time $|T| =0$. We might
assume that the particle has an interaction at spatial position x.
Then we might expect that having arrived at position x we must
consider each of the temporally precedent spatial positions where
an interaction did NOT occur: NOT(x-1), NOT(x-2)... where there
could have been but was no interaction. For a probability
distribution that was identically and uniformly distributed this
would be straightforward - we could examine the n positions  -
each separated by a distance $\Delta x = d $  prior to the
interaction at x:\\

\ldots\ldots\ldots\ldots*\\[-4mm]
 \hspace* {25mm}$ x = n\Delta x$
\\

We can define the probability of occurrence in a very short space
$\Delta x$ as $(B\Delta x)$ where B is the probability density.

So the probability of non-occurrence at a very short space is
$(1-B\Delta x)$

If a distance x is travelled before an interaction then where P(x)
is the probability density:

$P(x)\Delta x = (1-B\Delta x)^{n} B\Delta x$

 $P(x) = B(1-B\Delta x)^{n}$

For a large x then $n=\frac{x}{\Delta x}\longrightarrow\infty$

i.e. we might at first expect:

\beq  P(x) = B e^{-Bx} \eeq

However, B, the probability density of an interaction at each
position varies according to the number of alternative energy
positions at each possible spatial position x. As we look more
carefully at the energy position alternatives at each spatial
position x we see that the range of possible \textit{energy
positions itself} will vary at different spatial positions. We
therefore have to examine the probability density of position in a
very short space as a variable which depends on energy levels and
which we will label $B(rt')$. We are using the probability of
energy position as the probability density of spatial position at
a small point in space.

Let us assume that for each occasion that the IFE moves from one
energy position to another or from one spatial position to another
there is a primary uniform probability A of interaction for an IFE
with another (group of) IFEs (that depends on the state of the
other group of IFEs.

We might therefore assume that to arrive at the probability for an
interaction at a specific energy position (rt') at a spatial
position x we
sum all of the probabilities for each possible energy position at x (see Diagram 6 below).\\

Diagram 6

\hspace* {-11mm}$r_{x}t'$
\\[-4mm]
 \vdots\hspace* {7mm}*
\\[-4mm]
 \vdots\hspace* {8mm}.
\\[-4mm]
\vdots\hspace* {8mm}.
\\[-4mm]
\vdots\hspace* {8mm}.
\\[-4mm]
\vdots\hspace* {8mm}.
\\[-4mm]
\vdots\hspace* {8mm}.
\\[-4mm]
\vdots\hspace* {8mm}.
\\[-8mm]
\ldots\ldots\ldots\ldots\ldots\ldots \hspace* {8mm}\\[-4mm]\hspace* {10mm}x\\

Consider the probability $P(r_{x}t')$ for an interaction at a
single energy position $(rt')$ at spatial position x. We note that
in order for there to be an interaction at the energy position
$(rt')$, we must have had no interactions at each of the previous
possible (and temporally precedent) energy points (r-1)t',
(r-2)t'… etc.

To calculate the probability of an interaction at a particular
energy position we use a similar method to that initially assumed
for spatial position.

We define that the primary uniform probability of an interaction $
= A$

Now for a particular spatial position x the probability density of
having an interaction at one of the energy positions will be
inversely proportional to the time taken to move through the
potential number of energy positions at x $= A.\frac{1}{r_{x}t'}$
where $r_{x}$ is the number of possible energy positions at x.

i.e. the probability of interaction in a short interval of time $=
A \frac{1}{r_{x}t'}t'$

And the probability of non-occurrence in a short interval of time
is $(1 - A \frac{1}{(r_{x}t')} t')$

If an interval of time rt' passes before an interaction then where
$P(rt'/x)$ is the probability \textit{density} of (rt') at a given
x:

$ P(rt' / x)= A (1 - A \frac{1}{(r_{x}t')} t')^{r} = A ( 1-A (
\frac{r(t')}{r (r_{x}t')})^r$

For large (rt') then $(1-A\frac{rt'}{r (r_{x}t')})^r $ tends to
$e^{-A \frac{rt'}{r_{x}t'} }$

\beq P(rt'/x) = A e^{-A\frac{(rt')}{r_{x}t'}} \eeq

It is straightforward to calculate mean and variance using
this.\footnote{note that we can calculate the mean and variance as
follows:

   $$ Mean = k_o = <k>=\int^\infty_{-\infty}k(A)e^{-kA}dx  =\frac{A}{A^2} = \frac
   {1}A
   $$

$$   variance = \sigma^2_k=<k^2> - (<k>)^2$$

$$ <k^2> =\int^\infty_{-\infty}k^2(A)e^{-kA}dx = \frac{2A}{A^3} =
\frac {2}{A^2} $$

$$ i.e. \sigma^2_k=\frac {1}{A^2}$$}

However, we must consider not only a single given spatial position
at x (=nd), but further alternative possible \textit{spatial}
positions such as at x = (n-1)d, (n-2)d...etc.

Since the total time magnitude $|T|$ can be composed in more than
one way, then for a particular energy position (rt') we must
consequently consider not only non-occurrences at (r-1)t',
(r-2)t'..., but also for each of these energy positions, the non
occurrences at all the coterminous \textit{spatial} positions
which provide the same time magnitude $|T| = |rt'|$. In this case,
in order to establish $P(|T|)$ we must consider all the ways in
which it can be formed from the combination of the the first
spatial position, the second spatial position etc... It is
sometimes useful to highlight $P(|rt'|)= B (rt')$ when it can be
over more than one spatial position.

Consider firstly a second possible spatial position only. Given a
\textit{specific} spatial position then calculation of the
probability of an energy position (rt') requires us only to
consider all the possible alternative energy positions where no
interaction occurred at (r-1)t', (r-2)t'...etc at that given
spatial position. However, with an alternative possible spatial
position we must account also for all the possible energy
positions at the second spatial position which in combination with
the spatial-time t* (caused by the movement in spatial position)
can comprise the same time magnitude equal to $|rt'|$ in the first
spatial position. These possible positions for a certain energy
state (rt') at a specific position x (=nd) potentially exist only
for those combinations of spatial and energy positions which have
the same time magnitude $|T| = |rt'|$ such that $|T| =
\sqrt{(nt^{*})^2 + (npt' + rt')^2 } $ where r is the energy
position that can occur at any spatial position.

This requires us to account for the probability density of the
potential spatial position at x which itself accounts for the
permitted probability density $P(r_{x}t')$. This probability for a
potential spatial position $r_{x}$ is simply:

\beq P(r_{x}) =  \frac{1}{(nd)} d\eeq

Let us consider the temporal arc in more detail. We can see that
for each possible interaction at a specific spatial position x and
energy position (rt') we must consider all of the possible
interactions at energy and spatial positions on an arc associated
with it. We should consider firstly how many possible positions
are on this arc. Here we make use of a calculation originated by
Gauss for analysing a fundamental point lattice (see below).
\footnote{We essentially wish to know the number of
\textit{potential} positions on the temporal arc formed through
the time magnitude $|T| = \sqrt{(nt^{*})^2 + (npt' + rt')^2 }$.
Since t* and t' are finite numbers, and since n, p and r are
integers then there exist only a small subset of positions on the
temporal arc that can exist to form $|T|$. Since this can
effectively be represented as the root of a sum of two squares,
then we effectively want to estimate the number of lattice points
$C(|T|)$ on the circumference of a circle of radius $|T|$.

We can apply a theory of point lattices for determining the number
of possible lattice points \textit{in and on} a circle C($|T|$) of
radius $|T|$. If we consider the circle at the origin of a
fundamental point lattice with each lattice point as the centre of
a unit square with sides parallel to the axes t* and t', then we
can analyse the area of all the squares whose \textit{centres} are
inside or on $C(|T|)$. This area $L(|T|)$ will comprise a number
of complete squares entirely within the circle, but also \textit{a
number of squares that are divided by the circle of radius $|T|$}

Some parts of squares with \textit{centres} inside the circle of
radius $|T|$ will remain outside of the circumference, and equally
there are some squares with \textit{centres} outside the circle
whose boundaries fit partly within the circle's perimeter. If we
theoretically shade in all the complete squares whose centres are
in or on the circle, then we can bound the shaded area $L(|T|)$
from below and above - we find the largest disk whose interior is
completely shaded, and the smallest disk whose exterior is
completely unshaded. Since the diagonal of a unit square is
$\sqrt{2}$ then all shaded squares must be contained in a circle
of radius $ = |T| + (\sqrt{2}/2)$. Similarly the circle whose
radius $ = |T| - (\sqrt{2}/2)$ is contained entirely within the
shaded squares.

Consequently

$ \pi(|T|^2 -\sqrt{2}|T| - \frac{1}{2}) \leq \pi(|T|^2
-\sqrt{2}|T| + \frac{1}{2}) \leq L(|T|) \leq \pi(|T|^2 +
\sqrt{2}|T| + \frac{1}{2}) $

Which implies that

$ |\frac{L(|T|)}{|T|^2} - \pi | \leq \pi (\sqrt{\frac{2}{|T|^2}}
+\frac{1}{2|T|^2}) $

Since $(\sqrt{\frac{2}{|T|^2}} +\frac{1}{2|T|^2})$ tends to 0 as $
|T|\rightarrow\infty$ then $ L(|T|)/|T|^2 \rightarrow\pi $

i.e. $L(|T|) = \pi |T|^2$.

This defines the number of lattice points both in and on a circle
of radius $|T|$. We require the number of points solely
\textit{on} the circle of radius $|T|$. Using elementary geometry
this is simply $C(|T|) = 2\pi|T|$.

Whilst the behaviour will be irregular in that different arcs will
have volatile numbers of potential compositions through nt* and
rt'(and some arcs will be effectively prime, composed through only
a single instance of n and rt') we can operate with an average
value for the number of possible positions on a variable temporal
arc, which will be effective if summed over a large/infinite
series - which is how we will be performing our summation of
probabilities. We therefore sum the first n values of $L(|T|)$
(the number of possible lattice positions on a circle of radius
$|T|$) and divide by n to obtain an associated average for the
total number of of ways for combining the two axes of time to form
the single time magnitude:

$ \frac{C(|T|)}{|T|} = \frac{C(0)+ C(1) + C(2) +...+C(|T|)}{|T|} $

We can therefore utilise $C(|T|) = 2\pi|T|$.}

This shows that $C(|T|)$ the number of permissable points on a
temporal arc that can compose a time magnitude $|T|$ is:

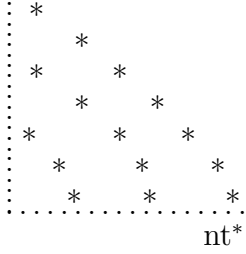
\beq C(|T|) = 2\pi |T| \eeq

If we examine the temporal arc closely, we see that to calculate
the probability of a particular energy (rt') we need to account
not only for all of the potential interactions that did
\textit{not} occur at energy positions (r-1)t', (r-2)t'... but
also for all of the feasible interactions that could have, but did
not occur at energy positions such as $(|rt'-t^*|),(|rt'-t^*
-1|)...$ at a second \textit{spatial} position - and further
$(|rt'-nt^*|),(|rt'-nt^* -1|)...$ at the nth spatial position.

\textbf{\textit{Calculation of the probability of NON-interactions
- represented by NOT (...) - requires us to sum the area of the
arc of every possible position at every possible spatial
position}}

Diagram 7\\
\hspace* {-6mm} qt'
\\[-4mm]
 \vdots\hspace* {2mm}*
\\[-4mm]
 \vdots\hspace* {8mm}*
\\[-4mm]
\vdots\hspace* {2mm}*\hspace* {9mm}*
\\[-4mm]
\vdots\hspace* {8mm}*\hspace* {8mm}*
\\[-4mm]
\vdots\hspace* {1mm}*\hspace* {10mm}*\hspace* {7mm}*
\\[-4mm]
\vdots\hspace* {5mm}*\hspace* {9mm}*\hspace* {8mm}*
\\[-4mm]
\vdots\hspace* {7mm}*\hspace* {8mm}*\hspace* {9mm}*
\\[-8mm]
\ldots\ldots\ldots\ldots\ldots\ldots \hspace* {23mm}
\\[-4mm]
\hspace* {25mm} nt$^{*}$\\

The mechanics for this calculation are facilitated if we work
backwards and investigate historically the non-occurrences of
interactions for spatial and energy positions.

To illustrate this technique, let us consider a simplified
example. For this example, we shall imagine the probability of NOT
having an interaction at an energy magnitude of $3$ where there
are only 3 possible spatial positions and 3 possible energy
positions. In order to (over)simplify this example further, we
shall also assume that each spatial position involves a
straightforward addition of a single value $t^* = t' = 1$.

Then we arrive at a layered iteration of probabilities that form a
NON-interaction :

\hspace* {-20mm} $ P( \textnormal {NOT} |(rt')|3)= NOT [
(x_{0}3)(x_{1}0)(x_{2}0) + (x_{0}0)(x_{1}2)(x_{2}0) +
(x_{0}0)(x_{1}0)(x_{2}1) + (x_{0}1)(x_{1}1)(x_{2}0)]$
\\[-4mm]
\hspace* {33mm} $\vdots $\hspace* {38mm} $\vdots $
\\[-4mm]
\hspace* {33mm}$ \textnormal {NOT}(|(rt')|2)...$\hspace* {12mm}$
\textnormal {NOT}(|(rt')|1)$

\hspace* {-20mm} $P( \textnormal {NOT} (|(rt')|2) = NOT [
(x_{0}2)(x_{1}0)(x_{2}0) + (x_{0}0)(x_{1}1)(x_{2}0)]$
\\[-4mm]
\hspace* {35mm} $\vdots $\hspace* {35mm}
\\[-4mm]
\hspace* {33mm}$ \textnormal {NOT}(|(rt')|1)$

\hspace* {-20mm} $P( \textnormal {NOT} (|(rt')|1) = NOT [
(x_{0}1)(x_{1}0)(x_{2}0)] $

It is not surprising that matrices have been applied in quantum
mechanics; however we shall not analyse this method further in
this paper.

There are two important points to note here. Firstly, because we
work backwards, we are investigating non-occurrences of
interactions and this means that instances such as
$(x_{0}1)(x_{1}1)(x_{2}0)$ must be considered probabilistically -
even though there is no such single possibility - i.e. the first
$(x_{0}1)$ is an instance of something that did not occur
\textit{in the history} of the second $(x_{1}0)$.

Secondly, for the $\textnormal {NOT} |(rt')|3)$ we include all
four possible groupings as well as all the $\textnormal {NOT}
|(rt')|2)$ non-happenings (and all of the $\textnormal {NOT}
|(rt')|1)$ in turn).

In order to simplify our calculation, we shall first make use of a
symbol $k$ to combine from equations 17,18 and 19:

$ k= \frac{2\pi|T|}{(n)(rt')}$

It will be noticed, interestingly, that $ k= \frac{2\pi}{\lambda}$

We note that through the symmetrical character of the squared time
magnitude there will be the same number of \textit{available}
energy and spatial positions.

We can use an efficient summation method which enables us to
aggregate all the possible probabilities. To illustrate this, we
can first calculate notionally for two spatial positions only -
(i.e. provided that there are only 2 energy positions)

$$ P(rt') = B(rt') = B(k) = \int^\infty_{-\infty}A e^{-Ak'} . A e ^{A(k-k')}dk'  $$

With the constraint that k' and (k-k') are not negative - i.e.
both $A e^{-Ak'}$ and $A e ^{A(k-k')}$ are effectively Heaviside
step functions which we can represent with the addition of H(k')
and H(k-k'):

$$ \int^\infty_{-\infty}A e^{-Ak'}H(k') . A e ^{A(k-k')}H(k-k')dk'  $$

$$ = \int^k_0 A e^{-Ak'} . A e ^{A(k-k')}dk'  $$

$$ = A^2 ke^{-Ak}$$

Similarly for 3 positions we have :

$$ \int^\infty_{-\infty}A^2 ke^{-Ak}H(k') . A e ^{A(k-k')}H(k-k')dk'  $$

$$ = \int^k_0 A^2k e^{-Ak'} . A e ^{A(k-k')}dk'  $$

$$ = \frac{A^3}{2} k^2e^{-Ak}$$

and for all the possible n positions across the temporal arc we
can see through inference that we obtain:

\beq P(k) = \frac{A^n k^{n-1}}{(n-1)!}e^{-Ak} \eeq

Note that:
 \beq Mean = k_o = <k>=\int^\infty_{-\infty}\frac{(kA)^n}{(n-1)!}e^{-kA}dx  =\frac{n}{A}  \eeq

\beq   variance = \sigma^2_k = \frac {n}{A^2} \eeq

let a = (Ak)/n

Then $$P(k) = \frac {A n^{n-1}} {(n-1)!} a^{n-1}e^{-na}$$

If we replace with e = a-1 then

$$ P(k) = \frac {An^n} {n!} (1+e)^{n-1}e^{n(1+e)}$$

Assuming that n is large, we can express P(k) in a more convenient
manner using Sterling's factorial expansion:\footnote{Jeffreys}

 $$ n! = \sqrt{(2{\pi}n)}n e^{-n}$$

Then $$P(k) = \frac {A}{\sqrt{2{\pi}n}}(1+e)^{n-1}e^{-ne}$$

But $e^{-ne} = 1 -\frac{ne}{1!}+\frac{(ne)^2}{2!}...$

and  $(1+e)^{n-1}=1+(n-1)e+\frac{(n-1)(n-2)e^2}{2}+...$

Then $$ P(k) =
\frac{A}{\sqrt{2{\pi}n}}(1-e-\frac{1}{2}(n-2)e^2+...)$$

$$ = \frac{A}{\sqrt{2{\pi}n}}e^{\frac{1}{2}n(e-1)^2}$$

$$ =\frac{A}{\sqrt{2{\pi}n}}e^{-\frac{1}{2}n(a-1)^2} $$

substituting back for a = (Ak)/n

$$=\frac{A}{\sqrt{2{\pi}n}}e^{-\frac{1}{2}n (\frac{Ak}{n}-1)^2}$$

$$ =\frac{A}{\sqrt{2{\pi}n}}e^{-\frac{1}{2}\frac{(Ak-n)^2}{n}} $$

$$=\frac{A}{\sqrt{2{\pi}n}}e^{-\frac{1}{2}
\frac{(k-\frac{n}{A})^2}{\frac{n}{A^2}}}$$

But from (19) and (20) $k_0 = \frac{n}{A}$ and $\sigma^2 =
\frac{n}{A^2}$

Then

\beq P(k) =\frac{1}{\sqrt{2{\pi}}\sigma_k}e^{-\frac{1}{2}
\frac{(k-k_0)^2}{(\sigma_k)^2}}\eeq

Since P(k) is an expression of k and since $k= \frac{2\pi
|T|}{\imath (nd)(rt')} $ then we note that $P(k^*) = P(-k)$

This provides an expression for the probability of a specific
interaction at a specific energy position but it does not account
for the spatial location.

To calculate P(x) the probability of an interaction at a specific
spatial position x, we sum all of the alternative P(k)'s at any
given x and ensure that we allow for every preceding non-event at
(x-1),(x-2)...

We know that the probability of NOT having an interaction at
spatial position x is given by $e^{-ux}$.

We must consider each P(k) at a given spatial position x, over
every probability density for each position of x whilst
eliminating every other (NOT x) position - along with the sum of
every feasible P(k) at each of these positions:

\beq P(x) = \int^\infty_{-\infty}P(k)
\{\int^\infty_{-\infty}P(u-k).e^{-ux}du\} dk \eeq

Now since P*(k)= P(-k) then:

$$ P(x) =\int^\infty_{-\infty}
P(k){\int^\infty_{-\infty}P^*(k-u).e^{-ux}du} dk $$

Using the notation for a Fourier Transform where the Fourier
Transform of P(k) is: $FT(P(k))
=\int^\infty_{-\infty}P(k)e^{-\imath kx}dk$ we can show that:

$$\int^\infty_{-\infty}
P^*(k-u)e^{-ux}du =
\int^\infty_{-\infty}P^*(k-u).e^{-(k-u)x}e^{-ux}d(k-u) =e^{-ux}
FT(P^* (k))
$$

$$ \textnormal {Hence } P(x) =\int^\infty_{-\infty} P(k) e^{-kx}FT(P^*(k))dk $$

$$ = FT (P^*(k)) . FT (P(k)) $$

Then \beq P(x) = |FT (P(k))|^2 \eeq

\textbf{That is the probability of finding an IFE at position x is
the square of the magnitude of the Fourier transform of P(k). This
is the probabilistic heart of quantum mechanics.}

We can consequently define a function $\psi(x)$ - we will name it
``the probability function" - such that $\psi (x)$ = $FT (P(k))$
and thus:

\beq P(x) = |\psi (x)|^2  \eeq

Since from (21)$$ P(k)
=\frac{1}{\sqrt{2{\pi}}\sigma_k}e^{-\frac{1}{2}
\frac{(k-k_0)^2}{(\sigma_k)^2}}$$

Then $$ P(x) =
|\int^\infty_{-\infty}\frac{1}{\sqrt{2{\pi}}\sigma_k}e^{-\frac{1}{2}
\frac{(k-k_0)^2}{(\sigma_k)^2}}e^{-\imath kx}dk|^2 $$

To assist in calculation we can use $k' = k - k_0$

$$ P(x) = |
\frac{1}{\sqrt{2\pi}\sigma_k}\int^\infty_{-\infty}e^{\frac{-\frac{1}{2}(k')^2}{\sigma_k^2}}e^{\imath(k'+k_0)x}
 dk'|^2
$$

$$= |
\frac{e^{\imath
k_{0}x}}{\sqrt{2\pi}\sigma_k}\int^\infty_{-\infty}e^{(\frac{-\frac{1}{2}(k')^2}{\sigma_k^2}
+\imath k'x)}dk'|^2
$$

$$= |
\frac{e^{\imath
k_{0}x}}{\sqrt{2\pi}\sigma_k}\int^\infty_{-\infty}e^{\frac{1}{2\sigma_{k}^2}(\imath
k' - \sigma_{k}^2 x)^2} .e^{\frac{1}{2}\sigma_k^2x^2}dk' |^2
$$

Since, from integration tables: $\int^\infty_{-\infty}e^{-x^2}dx =
\sqrt{\frac{\pi}{x}}$ (where x can be complex):

$$ P(x) = |\frac{e^{\imath
k_{0}x}}{\sqrt{2\pi}\sigma_k}e^{\frac{1}{2}\sigma_k^2x^2}\sqrt{\pi
2 \sigma_k^2 } |^2 $$

$$ i.e. P(x) = 2\pi\sigma_k^2e^{\sigma_k^2x^2} $$

Thus P(x) for a free particle also has a Gaussian distribution. A
number of significant consequences derive from this explanation.

Although we can derive and make use of the ``probability function"
$\psi(x)= FT(P(k))$, it is a strange creature of mixed real and
imaginary heritage - and it lurks like a half human half bull
Minotaur in a labyrinth of misunderstanding from which reason
never escapes. It has only mathematical significance and no direct
reference.

We can, nevertheless, note an interesting feature of the interplay
between P(k) and $\psi(x) = FT(P(k))$.

Firstly, we note a property of the differential of P(x) which we
can indicate by P'(x):

$$\int^\infty_{-\infty}P'(x)e^{-ikx}dx = e^{-ikx}P(x)|^\infty_{-\infty} +ik\int^\infty_{-\infty}P(x)e^{-ikx}dx$$

\beq \int^\infty_{-\infty}P'(x)e^{-ikx}dx = ikFT(P(k))\eeq

Secondly we note that for a combination of such probability
functions: say P(k) and another similar probability function in k
Q(k) then as probability functions are not negative then provided
that a factor $\alpha$ is non-zero (where $P^*(k)$ indicates the
complex conjugate of P(k)) :

$$\int^\infty_{-\infty}\{P(k) + \alpha Q(k)\}^2dk>0$$

Then
$$\int^\infty_{-\infty}|P(k)|^2 dk +
\alpha^2\int^\infty_{-\infty}|Q(k)|^2 dk +
2\alpha\int^\infty_{-\infty}\{P(k)Q(k)\}dk > 0$$

We can solve this as a quadratic equation for $\alpha$ and hence:

$$ \{2\int^\infty_{-\infty}P(k)Q(k)dk \}^2 -
4\int^\infty_{-\infty}|P(k)|^2dk \int^\infty_{-\infty}|Q(k)|^2 dk
\leq0 $$

Then \beq \{\int^\infty_{-\infty}P(k)Q(k)dk \}^2 \leq
\int^\infty_{-\infty}|P(k)|^2dk \int^\infty_{-\infty}|Q(k)|^2dk
\eeq

Using basic definitions for the variance of x and k, we form the
multiple:

$$ \sigma_x^2\sigma_k^2 = \int^\infty_{-\infty}x^2|P(x)|^2dx.
\int^\infty_{-\infty}k^2 |(P(k))|^2dk $$

However, we can show that:\footnote {This will be familiar as the
proof of the Parseval identity:

$$\int^\infty_{-\infty}P(k) P^* (k) dk = \int^\infty_{-\infty}
P(k)\{\int^\infty_{-\infty}FT(P^*(k))e^{rk}\}dk$$

$$ =\int^\infty_{-\infty}\int^\infty_{-\infty}P(k)e^{rk} P^*
(k)dk$$

$$ =\int^\infty_{-\infty} FT(P(k))
FT^*(P(k))$$

}

 $$\int^\infty_{-\infty}|(P(k))|^2dk =
\int^\infty_{-\infty}|FT(P(k))|^2dk$$

$$ \textnormal{Hence } \sigma_x^2\sigma_k^2=
\int^\infty_{-\infty}|xP(x)|^2dx.\int^\infty_{-\infty}|ikFT(P(k)|^2dk$$

From (25):

$$ =
\int^\infty_{-\infty}|xP(x)|^2dx.\int^\infty_{-\infty}|P'(x)|^2dx$$

From (26):

$$ \sigma_x^2\sigma_k^2\geq\int^\infty_{-\infty}|x(P(x) P^{*'}(x)dx|^2 $$

$$ \sigma_x^2\sigma_k^2\geq\int^\infty_{-\infty}|x\frac{d}{dx}|(P(x))|^2dx|^2 $$

$$ \sigma_x^2\sigma_k^2\geq\frac{1}{4}\int^\infty_{-\infty}{|P(x)^{2}dx|}^{2} $$

And since $\int^\infty_{-\infty}|P(x)^2dx|^2$ is the probability
of finding the IFE \textit{anywhere} =1

Then

$$ \sigma_x^2\sigma_k^2 \geq\frac{1}{4} $$

\beq \sigma_x\sigma_k \geq\frac{1}{2} \eeq

\section{Momentum and the Concept of Mass}

It is useful to consider the apparent combined velocity of an IFE
which is moving with a velocity $v = \frac{d}{\sqrt{(t^* )^2 +
(pt')^2 }}$ away from a notional fixed reference point and another
IFE which is moving away in the other direction from the fixed
reference point at a velocity $u = \frac{d}{\sqrt{(t^* )^2 +
(qt')^2 }}$. This effectively becomes a method for perceiving the
resultant velocity of two velocities added together.

We shall consider what occurs in a time $\sqrt{(t^* )^2 +
(pt')^2}$ :

The distance D travelled in this time is:

$ D = d + \frac{d}{\sqrt{(t^* )^2 + (qt')^2 }}\sqrt{(t^* )^2 +
(pt')^2}$

However, in analysing the amount of \textit{time} we should employ
in formulating the combined velocity of both IFE's, there is a
further complication here. During the period of time $\sqrt{(t^*
)^2 + (pt')^2}$ which accounts for a movement in space d for the
first IFE, then we have to account for an additional number of
\textit{spatial-time} points that would have been covered by the
second IFE (determined by the trigger-point qt').

In order to establish how many ``extra" incidents of spatial-time
t* there are in this time, we can consider that in a theoretical
amount of time stretching across $\sqrt{(t^* )^2 +
(pt')^2}\sqrt{(t^* )^2 + (qt')^2 }$ we will have an extra number N
of incidents of $t^*$ where:

$ N = \sqrt{(\sqrt{(t^* )^2 + (pt')^2} + \sqrt{(t^* )^2 +
(qt')^2})^2 - \{(\sqrt{(t^* )^2 + (pt')^2})^2 + (qt')^2}\} $

This gives us a \textit{rate} of discrepancy of extra $t^*$ per
unit of time such that:

$rate = \frac{\sqrt{(\sqrt{(t^* )^2 + (pt')^2} + \sqrt{(t^* )^2 +
(qt')^2})^2 - \{(\sqrt{(t^* )^2 + (pt')^2})^2 + (qt')^2} \}
}{\sqrt{(t^* )^2 + (pt')^2}\sqrt{(t^* )^2 + (qt')^2 }}$

In an amount of time $\sqrt{(t^* )^2 + (pt')^2}$ there will be
$\frac{\sqrt{(t^* )^2 + (pt')^2}}{\sqrt{(t^* )^2 + (qt')^2 }}$
opportunities for an extra ``skip" of space-time.

The total number of extra incidents of $t^*$ will be:

 $\frac{\sqrt{(t^* )^2 + (pt')^2}}{\sqrt{(t^* )^2 + (qt')^2
}}\frac{\sqrt{(\sqrt{(t^* )^2 + (pt')^2} + \sqrt{(t^* )^2 +
(qt')^2})^2 - \{(\sqrt{(t^* )^2 + (pt')^2})^2 +
(qt')^2}\}}{\sqrt{(t^* )^2 + (pt')^2}\sqrt{(t^* )^2 + (qt')^2 }}$

Then the amount of time t we have to consider when calculating the
combined velocity of the two IFEs is:

$t = \sqrt{(\sqrt{(t^*)^2 + (pt')^2})^2 + (t^*)^2
[\frac{\sqrt{(t^* )^2 + (pt')^2}}{\sqrt{(t^* )^2 + (qt')^2
}}\frac{\sqrt{(\sqrt{(t^* )^2 + (pt')^2} + \sqrt{(t^* )^2 +
(qt')^2})^2 - \{(\sqrt{(t^* )^2 + (pt')^2})^2 +
(qt')^2}\}}{\sqrt{(t^* )^2 + (pt')^2}\sqrt{(t^* )^2 + (qt')^2
}}]}$

$ = \sqrt{(\sqrt{(t^*)^2 + (pt')^2})^2 + (t^*)^2
[\frac{2(\sqrt{(t^* )^2 + (pt')^2}}{\sqrt{(t^* )^2 + (qt')^2 }} +
\frac{(t^*)^2}{(t^*)^2 + (qt')^2}} ]$

$ = \sqrt{(t^*)^2 + (pt')^2} + \frac{(t^*)^2}{\sqrt{(t^* )^2 +
(qt')^2 }} + 2(t^*)^2 \frac{\sqrt{(t^* )^2 + (pt')^2}}{\sqrt{(t^*
)^2 + (qt')^2}} $

$ = [ \sqrt{\sqrt{(t^*)^2 + (pt')^2} + \frac{(t^*)^2}{\sqrt{(t^*
)^2 + (qt')^2}}} ]^2$

$= \sqrt{(t^* )^2 + (pt')^2} + \frac{(t^*)^2}{\sqrt{(t^* )^2 +
(qt')^2}} $

Then the combined velocity V of the two IFE's is:

\beq  v = \frac{ d + \frac{d}{\sqrt{(t^* )^2 + (qt')^2
}}\sqrt{(t^* )^2 + (pt')^2}}{\sqrt{(t^* )^2 + (pt')^2} +
\frac{(t^*)^2}{\sqrt{(t^* )^2 + (qt')^2}}} \eeq

We can now consider a particular case of interest. Consider two
IFE's of equal rest mass $m_{0}$ and equal velocity $u$ colliding
in a non-elastic way from opposite directions (say a mass moving
from the left and a mass moving from the right), resulting in a
stationary object of mass $M_{0}$.

We will suppose that mass is not necessarily fixed and hypothesise
that the rest mass varies so that the mass $m(v)$ may be different
from the mass $m_{0}$ when stationary.

We can also imagine that in this case we sit on the second IFE
mass moving from the right. From this perspective the mass moving
from the left has an effective combined velocity V and has a mass
m(V). It then hits the IFE (on which we sit) of mass $m_{0}$ which
results in an IFE of mass $m(u)$ moving with a velocity u.

Effectively velocity V is the combined velocity of two equal
velocities each moving with velocity u. We can see from equation
$30$ above that the effective velocity of two combined equal
velocities $u= \frac{d}{\sqrt{(t^* )^2 + (pt')^2}}$ is:

\beq V = \frac{2d\sqrt{(t^* )^2 + (pt')^2}}{2(t^* )^2 + (pt' )^2}
\eeq

We will now employ two fundamental laws which, I have to say,
appear initially as quite grand assumptions. These are:

(1) Conservation of Momentum

 i.e. $m(V) V = m(u)u$

 (2) Conservation of Mass

 i.e. $m(V) + m_{0} = M(u)$

 It will be seen later that
 these two laws are both concerned with the same fundamental which is the
 conservation of time.

 If we combine these two conservation laws and eliminate $m(u)$, we obtain:

 \beq \frac{m{V}}{m{0}}= \frac{u}{V-u} \eeq

 Making use of $u= \frac{d}{\sqrt{(t^* )^2 + (pt')^2}}$ and equation 30, we obtain:

$$ \frac{m(V)}{m_{0}} =  \frac{\frac{d}{\sqrt{(t^* )^2 +
(qt')^2}}}{\frac{2d\sqrt{(t^* )^2 + (qt')^2}}{2(t^* )^2 + (pt'
)^2} -\frac{d}{\sqrt{(t^* )^2 + (qt')^2}}}= \frac{d}{\frac{2d(t^*
)^2 + (qt')^2}{2(t^* )^2 + (pt' )^2} -d}$$

\beq \textnormal{Then }\frac{m(V)}{m_{0}}= \frac{2(t^*)^2 +
(pt')^2}{(pt')^2} \eeq

It is instructive to note that if we multiply by $
\frac{d^2}{(t^*)^2}$:

\beq m(V)\frac{d^2}{(t^*)^2} = m_{0}\frac{d^2}{(t^*)^2} +
2m_{0}\frac{d^2}{(pt')^2} \eeq

The second expression on the right indicates a multiple of the
rest mass with some form of the square of the velocity.

If we compare with a traditional $\frac{1}{2}m_{0}V^2$ Newtonian
formulation of kinetic energy, then applying $V =
\frac{2d\sqrt{(t^* )^2 + (pt')^2}}{2(t^* )^2 + (pt' )^2}$:

 $$2m_{0}\frac{d^2}{(pt')^2} - \frac{1}{2}m_{0}V^2 = 2m_{0}\frac{d^2}{(pt')^2} - \frac{2m_{0}d^2\{(t^*)^2 + (pt')^2\}}{\{2(t^*)^2 +
(pt')^2\}^2}$$

$$ = \frac{(t^*)^2}{(pt')^2}\frac{2(t^*)^2}{\{2(t^*)^2 +
(pt')^2\}^2}$$

This last term becomes negligible when $p \gg t^*$ i.e. for speeds
much less than the speed of light, and hence for such speeds
$2m_{0}\frac{d^2}{(pt')^2}\sim \frac{1}{2}m_{0}V^2$. This
indicates that equation $34$ is a relation for the \textit{energy}
of the IFE, where the term $2m_{0}\frac{d^2}{(pt')^2}$ indicates
the kinetic energy of the particle and consequently we have a term
for the rest energy of the IFE:

$$ E_{0} = m_{0}\frac{d^2}{(t^*)^2} $$

And for the total energy of the IFE:

\beq E_{T} = m(V) \frac{d^2}{(t^*)^2} \eeq

 These are, of course, instances of Einstein's familiar expression
$E = mc^2$.

We can further note that:

$$ m(V) V = m_{0} \{\frac{2(t^*)^2 +
(pt')^2}{(pt')^2}\} \{\frac{2d\sqrt{(t^* )^2 + (pt')^2}}{2(t^* )^2
+ (pt' )^2}\}$$

$$ m(V) V = \frac{2dm_{0}\sqrt{(t^* )^2 +
(pt')^2}}{(pt')^2}$$

From (15) we know that $E = \frac{h}{(rt')}$ and from equation
(35) $m_{0}=\frac{h (t^*)^2}{(rt')d^2}$

$$\textnormal{Then } m(V)V = \frac{h 2(t^*)^2 \sqrt{(t^* )^2 +
(pt')^2}}{d (rt')(pt')^2} =  \frac{h \sqrt{(t^* )^2 + (pt')^2}}{d
(rt') } \frac{2 (t^*)^2}{(pt')^2}$$

Yet this is an expression concerning the wavelength of the
combined IFE which moves with speed $u = \frac{d}{\sqrt{(t^* )^2 +
(pt')^2}}$ which indicates that:

\beq m(V)V = \frac{h}{\lambda} \eeq

Since $ p = \frac{h}{\lambda} = \frac{hk}{2\pi}$ then

\beq \sigma_x\sigma_p \geq\frac{h}{4\pi}\eeq

This is the familiar expression of Heisenberg's Uncertainty
Principle.

\section{Conclusions}

The explanations covered in this paper underpin the theories of
relativity and quantum physics - and provide clarifying reasons
for some confusing aspects of these theories - including why the
speed of light has a maximum, why mass should be connected to
energy, the perceived differences in experienced time for moving
and stationary entities, how the concepts for the speed of light c
and Planck's constant relate more fundamentally to the units of
space-time t* and energy-time t', and the quantum uncertainty of
position and momentum. A later paper is intended on the theory of
gravity.

\section{References}

Jeffreys H: ``Theory of Probability" 1939, OUP.\\

Gauss C F: ``Werke" (Gottingen: Gesellschaft der Wissenschafften,
1863-1933)\\

Olds C D, Lax A, Davidoff G: ``The Geometry of Numbers" 2000, The
Mathematical Association of America

\end{document}